\documentclass[prb,aps,amsmath,amssymb,amsfonts,floatfix,superscriptaddress,showpacs,twocolumn]{revtex4}
\usepackage{graphicx}
\usepackage{color}
\usepackage{bbm}
\usepackage[utf8]{inputenc}
\usepackage{dcolumn}
\usepackage{hyperref}
\usepackage{textcomp}
\usepackage[english]{babel}

\hypersetup{colorlinks=true,breaklinks,linkcolor=blue,urlcolor=blue,citecolor=blue}
\usepackage[caption=false]{subfig}

\usepackage{comment}
\usepackage{tikz}
\usetikzlibrary{calc}
\usetikzlibrary{decorations.pathmorphing}
\usetikzlibrary{decorations.pathreplacing}
\usetikzlibrary{decorations.markings}

\tikzset{->-/.style={decoration={
  markings,
  mark=at position #1 with {\arrow{>}}},postaction={decorate}}}
\tikzset{-<-/.style={decoration={
  markings,
  mark=at position #1 with {\arrow{<}}},postaction={decorate}}}

\newcommand{\cpxi}{{\ensuremath{i}}}

\newcommand\teL{\mathrel{=\!\!\mathop:}}

\newcommand{\inu}{{\ensuremath{\cpxi\nu}}}
\newcommand{\kv}{\ensuremath{\mathbf{k}}}

\newcommand{\qv}{\ensuremath{\mathbf{q}}} 
 
\newcommand{\vc}[1]{\ensuremath{\mathbf{#1}}}

\newcommand{\av}[1]{\ensuremath{\left\langle #1 \right\rangle}}
\newcommand{\impav}[1]{\ensuremath{\left\langle #1 \right\rangle^{\text{imp}}}}

\newcommand{\up}{\ensuremath{\uparrow}}
\newcommand{\dn}{\ensuremath{\downarrow}}
\def \half {\ensuremath{\frac{1}{2}}}

\begin{document}

\title{Thermodynamic consistency of the charge response in dynamical mean-field based approaches}

\author{Erik G. C. P. van Loon}
\affiliation{Radboud University, Institute for Molecules and Materials, NL-6525 AJ Nijmegen, The Netherlands}

\author{Hartmut Hafermann}
\affiliation{Mathematical and Algorithmic Sciences Lab, France Research Center, Huawei Technologies Co. Ltd.}
\altaffiliation{Part of this work was conducted while the author was at Institut de Physique Th\'eorique (IPhT), CEA, CNRS, 91191 Gif-sur-Yvette, France.}

\author{Alexander I. Lichtenstein}
\affiliation{I. Institut f\"ur Theoretische Physik, Universit\"at Hamburg, Jungiusstra\ss e 9, D-20355 Hamburg, Germany}

\author{Mikhail I. Katsnelson}
\affiliation{Radboud University, Institute for Molecules and Materials, NL-6525 AJ Nijmegen, The Netherlands}

\date{\today}

\begin{abstract}
We consider the thermodynamic consistency of the charge response function in the (extended) Hubbard model.
In DMFT, thermodynamic consistency is preserved. We prove that the static, homogeneous DMFT susceptibility is consistent as long as vertex corrections obtained from the two-particle impurity correlation function are included.
In presence of a nonlocal interaction, the problem may be treated within extended DMFT (EDMFT), or its diagrammatic extension, the dual boson approach. We show that here, maintaining thermodynamic consistency requires knowledge of three- and four-particle impurity correlation functions, which are typically neglected. Nevertheless, the dual boson approximation to the response is remarkably close to consistency. This holds even when two-particle vertex corrections are neglected. EDMFT is consistent only in the strongly correlated regime and near half-filling, where the physics is predominantly local.
\end{abstract}


\maketitle

\section{Introduction}

Dynamical mean-field theory (DMFT)~\cite{Metzner89,Georges96,Kotliar06} has emerged as an important tool for the description of strongly correlated electron systems. The theory treats the local temporal correlations and can be applied at arbitrary interaction strength, allowing it to capture the Mott transition. Extended DMFT (EDMFT)~\cite{Si96,Kajueter96,Smith00,Chitra00,Chitra01} was developed to treat nonlocal interactions, that can lead to phenomena such as charge-ordering. Due to recent technical advances, EDMFT has received renewed attention~\cite{Ayral12,Ayral13,Hansmann13,Huang14} and has been adapted to spin systems (so-called spin-DMFT)~\cite{Otsuki13-2}. At the same time, diagrammatic extensions of DMFT and EDMFT have appeared, such as the dynamical vertex approximation (D$\Gamma $A)~\cite{Toschi07}, dual fermion~\cite{Rubtsov08,Rubtsov09}, the one-particle irreducible (1PI) approach~\cite{Rohringer13}, EDMFT+GW\cite{Sun02,Ayral12,Ayral13}, dual boson (DB)~\cite{Rubtsov12} and TRILEX~\cite{Ayral15}. 
These are advanced, nevertheless approximate methods to the correlated many-body problem. For any approximation it is desirable to fulfill certain requirements, such as the conservation laws of energy and charge~\cite{Baym61,Baym62} and thermodynamic consistency\cite{Baym61,Baym62}. While these cannot always be fulfilled in practice, it is nevertheless important to know for the interpretation of the results, if and how well these requirements are fulfilled.
In this work, we address the thermodynamic consistency of the linear charge response in DMFT, EDMFT and DB.

The response to an external perturbation is relevant for many experiments and for understanding the underlying physics. The linear response is related to correlation functions of the system via the celebrated Kubo formula~\cite{Kubo57}.
The correlation function is often more attractive to theorists. Especially for space- and time-dependent external perturbations  (e.g. fluctuating magnetic fields), the direct response may be difficult to calculate, because it involves the non-homogenous and non-equilibrium properties, whereas correlation functions of the unperturbed homogeneous equilibrium system are more accessible. In this case the Kubo formula provides a connection between the two and between theory and experiment.
The correspondence between the response and correlation functions is a property of the exact solution of the system and this equality is not automatically satisfied by approximate solutions~\cite{Dzyaloshinskii1976,Moriya1973,Moriya1985}. In fact, this kind of consideration was very important in the development of the theory of the electron gas~\cite{Geldart70}.

We call the linear response of an approximation thermodynamically consistent if the two ways, the direct calculation by varying the external field and that from the correlation function, yield the same result. 
The so-called zero- and one-particle quantities of DMFT are known to be thermodynamically consistent~\cite{Aichhorn06,Potthoff06}, i.e., the same density is obtained from deriving the grand potential and from the Green's function. The DMFT magnetic response~\cite{Brandt89} of the double-exchange model also turns out to be thermodynamically consistent~\cite{Fishman05},
while 
EDMFT and the related spin-DMFT~\cite{Otsuki13-2} appear to violate thermodynamic consistency. When \emph{both} the response and the correlation function can be calculated, thermodynamic consistency can be checked.

Here we apply such analysis to the charge susceptibility of an (extended) Hubbard model with local and nonlocal density-density interaction. We find that the charge response in DMFT can be obtained in a consistent manner provided vertex corrections from the impurity two-particle correlation function are included. In EDMFT and DB, thermodynamic consistency is in general violated and we show that the knowledge of the three- and four-particle impurity correlation functions would be required to restore this deficiency. 
While this is often impractical, the DB approximation turns out to contain the dominant contributions to yield a response that is consistent to very good approximation, over a broad interaction and filling range. We identify those contributions.

The paper is organized as follows:
The model and the Kubo formula for the charge response are introduced in Sec. \ref{sec:model}. In Sec. \ref{sec:local}, we examine the case of local interaction and show, by resorting to the DB approach, that DMFT is thermodynamically consistent in the Hubbard model. We then use thermodynamic consistency as a yardstick to measure how other approximations to the charge correlation function, namely DB without vertex corrections, approximation by the impurity susceptibility and the DMFT bubble, perform. In Sec. \ref{sec:nonlocal}, we apply the same approach to the extended Hubbard model and find that three- and four-particle interactions beyond a two-particle ladder summation are necessary to reach thermodynamic consistency. Finally we summarize our findings in Sec. \ref{sec:conclusion}. Additional technical facets and a detailed derivation of the DMFT response formula can be found in the Appendix.

\section{Model}
\label{sec:model}

We study a single-band, two-dimensional extended Hubbard model on a square lattice away from half-filling.
The Hamiltonian of the extended Hubbard model is given by
\begin{align}
  H &= -\sum_{\av{jk}\sigma} t_{jk} c_{j\sigma}^\dagger c_{k\sigma}^{\phantom{\dagger}} - \mu \sum_{j} n_j \notag\\
  &\phantom{=} + \sum_{j} U n_{j\up} n_{j\dn}  + \half \sum_{\av{jk}} V_{jk} n_j n_k, \label{eq:hmlt}
\end{align}
with $ n_j  = \sum_{\sigma} n_{j\sigma}  = \sum_{\sigma} c^\dagger_{j\sigma} c_{j\sigma}^{\phantom{*}}$, $\sigma=\up,\dn$ is the spin, $j$ and $k$ denote site indices and $c_{j\sigma}^\dagger$ ($c_{j\sigma}^{\phantom{\dagger}}$) are the creation (annihilation) operators for an electron with spin $\sigma$ on site $j$. In this work, we restrict ourselves to nearest-neighbor hopping and interaction, i.e., $t_{jk}=t$ and $V_{jk}=V$ when  $j$ and $k$ are nearest-neighbors and zero otherwise. The physical parameters are the hopping integral $t$, local interaction $U$, nonlocal interaction $V$ and chemical potential $\mu$. For our calculations, we take the half-bandwidth $D=4t$ as the energy unit and use inverse temperature $\beta=20$. 
As in EDMFT+GW~\cite{Sun02}, it is convenient for DB calculations to write the nonlocal interaction in Eq. \eqref{eq:hmlt} in terms of density fluctuations, which requires a Hartree shift $\mu^{\text{DB}} = \mu - \sum_{j\neq k} V_{jk} n_j$ in the chemical potential.

We study the dependence of the density $\av{n}$ on the chemical potential using the linear response formalism.
The Kubo formula relates the density response to a two-particle correlation function.
The electron compressibility $d \!\av{n}/d \mu$ is related to the static homogeneous correlation function as (Appendix~\ref{app:kubo}):
$
d\!\av{n}/d \mu  =- \lim_{\qv\to \vc{0}}\lim_{\omega\to 0}X_{\qv,\omega}
$.

\section{Local interaction}
\label{sec:local}

We first address the thermodynamic consistency for the simpler case of the Hubbard model ($V=0$) treated within DMFT~\cite{Georges96}. In DMFT, the lattice problem is mapped to an auxiliary single-site impurity problem with action 
\begin{align}
\label{simp:dmft}
S_{\text{imp}}[c^{*},c]=&-\sum_{\nu\sigma} c^{*}_{\nu\sigma}[\inu+\mu-\Delta_{\nu\sigma}]c_{\nu\sigma}\notag\\
&+ U\sum_{\omega}n_{\omega\up}n_{-\omega\dn}.
\end{align}
Here $\nu$ and $\omega$ are the fermionic and bosonic Matsubara frequencies respectively, $n_{\omega\sigma}=\sum_\nu c^{*}_{\nu\sigma} c_{\nu+\omega\sigma}$ and a normalization factor $\beta$ is implied in the sums over frequencies. The hybridization function $\Delta_{\nu\sigma}$ describes the electron hopping processes from and to the impurity and includes the effect of the lattice in a mean-field manner. It is determined self-consistently via the condition
\begin{align}
\sum_{\kv} G_{\kv\nu\sigma} = g_{\nu\sigma}, \label{eq:sc}
\end{align}
where $g_{\nu\sigma}$ is the Green's function of the impurity and $\sum_{\kv}$ denotes averaging over momenta. The lattice Green's function is given in momentum space as 
\begin{align}
 G^{-1}_{\kv\nu\sigma} = g^{-1}_{\nu\sigma} + \Delta_{\nu\sigma} - t_{\kv}, \label{eq:Gdmft}
\end{align}
where $t_{\kv}$ is the Fourier transform of the hopping.

\subsection{Density response}

In DMFT, the impurity density is the same as the lattice density as a direct consequence of the self-consistency condition. The impurity and lattice compressibilities, however, are different, since the impurity density depends on the chemical potential $\mu$ directly, but also via the hybridization function $\Delta_{\nu\sigma}$ which is determined self-consistently. As a consequence, the compressibility decomposes into the impurity compressibility and a part originating from the variation of $\Delta_{\nu\sigma}$ due to a change in chemical potential:
\begin{align}
 \frac{d\!\av{n}}{d \mu} &=  \frac{\partial\!\impav{n}}{\partial \mu} +  \sum_{\nu\sigma} \frac{\partial\! \impav{n}}{\partial \Delta_{\nu\sigma}} \frac{\partial \Delta_{\nu\sigma}}{\partial \mu}.\label{eq:dn/dm}
\end{align}
Here and in the following we use $\av{.}$ and $\impav{.}$ to denote lattice and impurity averages, respectively.
The right-hand side consists of three parts. 
The first is the impurity compressibility. The second part is proportional to a two-particle impurity correlation function $\partial\! \impav{n}/\partial\Delta_{\nu\sigma}\propto-\impav{n_{\omega=0}c^*_{\nu\sigma}c^{\phantom{*}}_{\nu\sigma}}+\impav{n}\impav{c^*_{\nu\sigma}c^{\phantom{*}}_{\nu\sigma}}$. According to Eqs. \eqref{eq:sc} and \eqref{eq:Gdmft}, the information about the lattice comes in through the variation of the self-consistent hybridization, i.e., $\partial \Delta_{\nu\sigma}/\partial \mu$. By working out the details, as done in Appendix \ref{app:dmftresponse}, the DMFT response can be expressed in terms of one- and two-particle impurity correlation functions. A crucial aspect in the derivation is that the impurity problem is solved (numerically) exactly using continuous-time quantum Monte Carlo~\cite{Rubtsov05,Werner06,Hafermann13}. The impurity response and correlation functions are therefore thermodynamically consistent. The final result is the static, homogeneous susceptibility of the DB approximation, which, for the Hubbard model, has been proven\footnote{The proof requires that the DB susceptibility is calculated with $\Delta_{\nu\sigma}$ chosen to satisfy the DMFT self-consistency condition \eqref{eq:sc}, and that there is no retarded impurity interaction [see Sec. \ref{sec:nonlocal}], i.e., $\Lambda_{\omega}=0$.} to be exactly equivalent to the  DMFT susceptibility including vertex corrections~\cite{Hafermann14-2}.

\begin{figure}
 \includegraphics[width=.9\columnwidth]{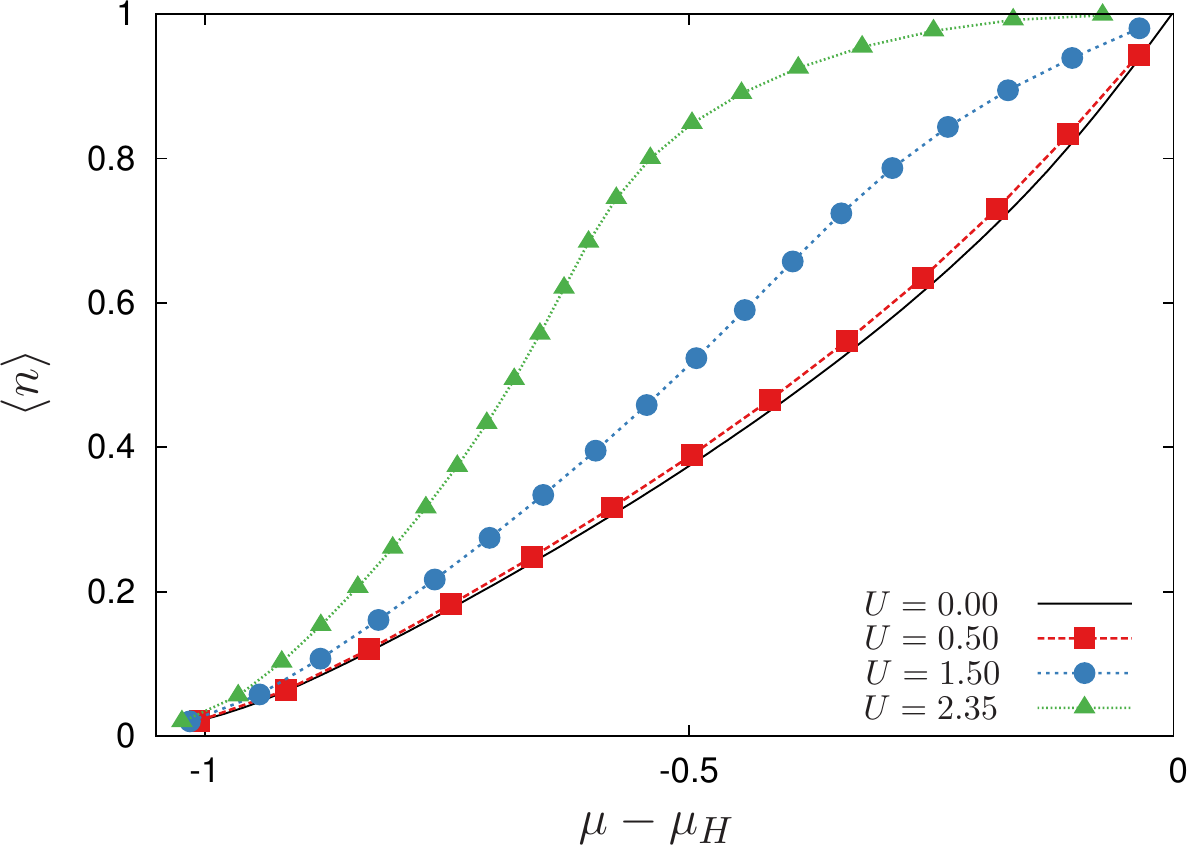}
 \caption{(Color online) Density $\av{n}$ obtained in DMFT as a function of chemical potential $\mu$ for different values of the local interaction $U$, at $t=0.25, T=1/20$. The Hartree contribution $\mu_H$ is subtracted from the chemical so that half-filling occurs at $\mu-\mu_H=0$ for all interaction strengths.
Close to the Mott transition ($U=2.35$) the chemical potential becomes flat near half-filling due to the development of a gap.}
  \label{fig:nmu}
\end{figure}

\begin{figure}
    \subfloat[$U=0.5$, $V=0$\label{fig:dndmu:U05}]{%
      \includegraphics[width=.9\columnwidth]{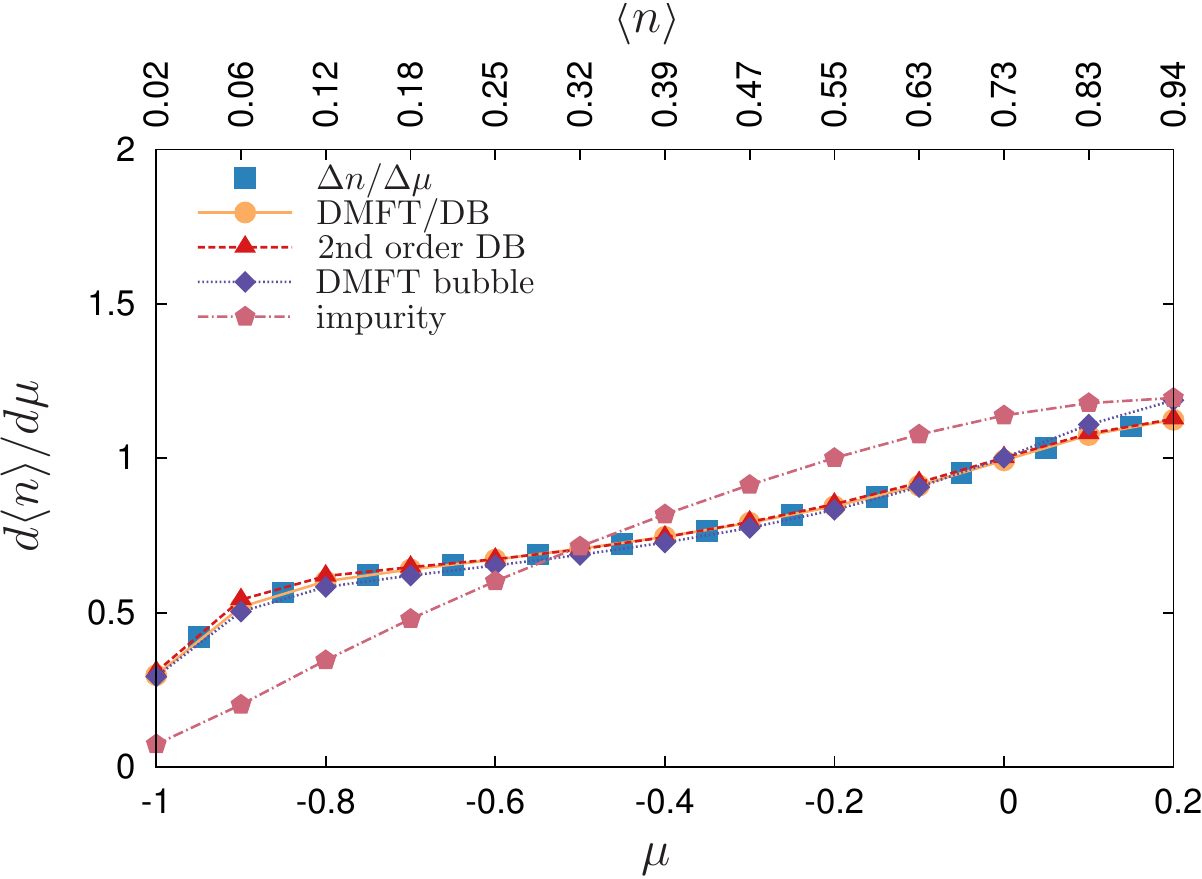}
    }
    \hfill
    \subfloat[$U=1.5$, $V=0$\label{fig:dndmu:U1500}]{%
      \includegraphics[width=.9\columnwidth]{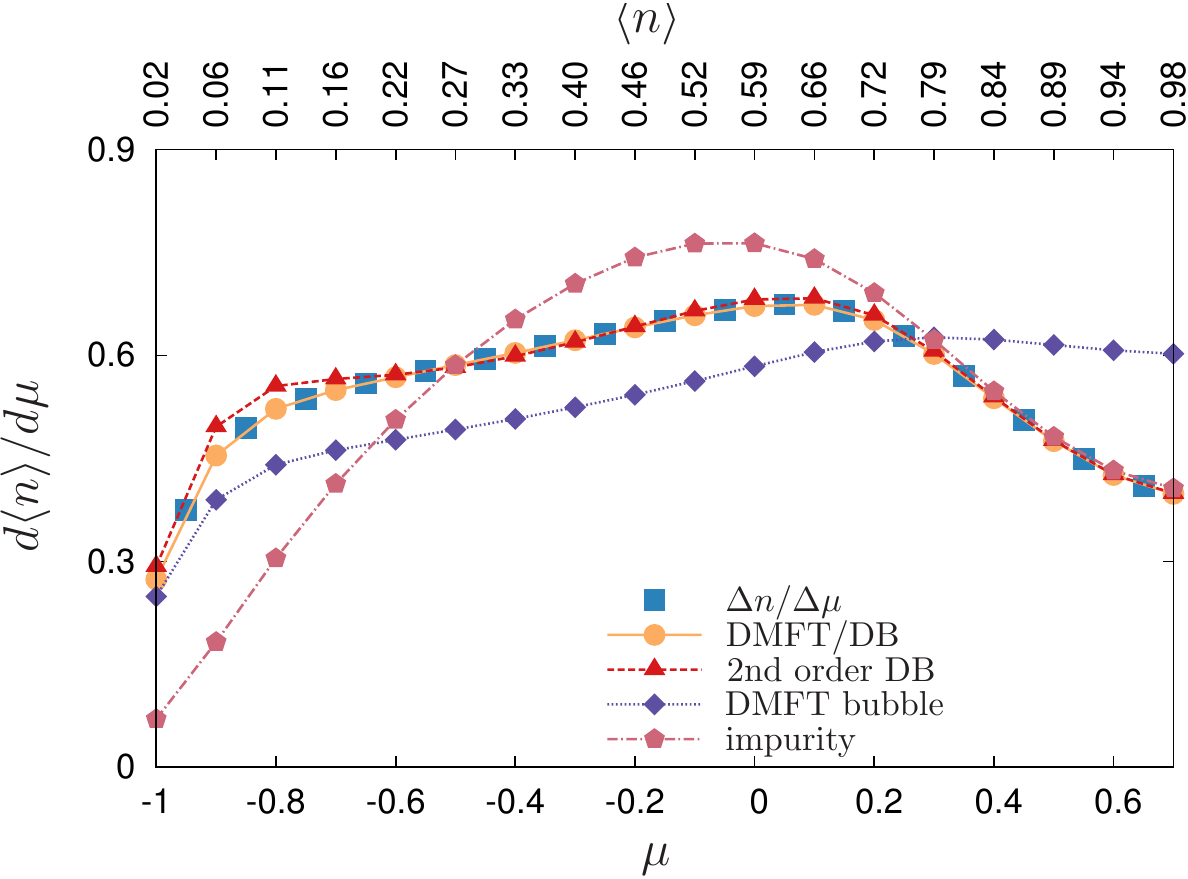}
    }
    \hfill
    \subfloat[$U=2.35$, $V=0$\label{fig:dndmu:U235}]{%
      \includegraphics[width=.9\columnwidth]{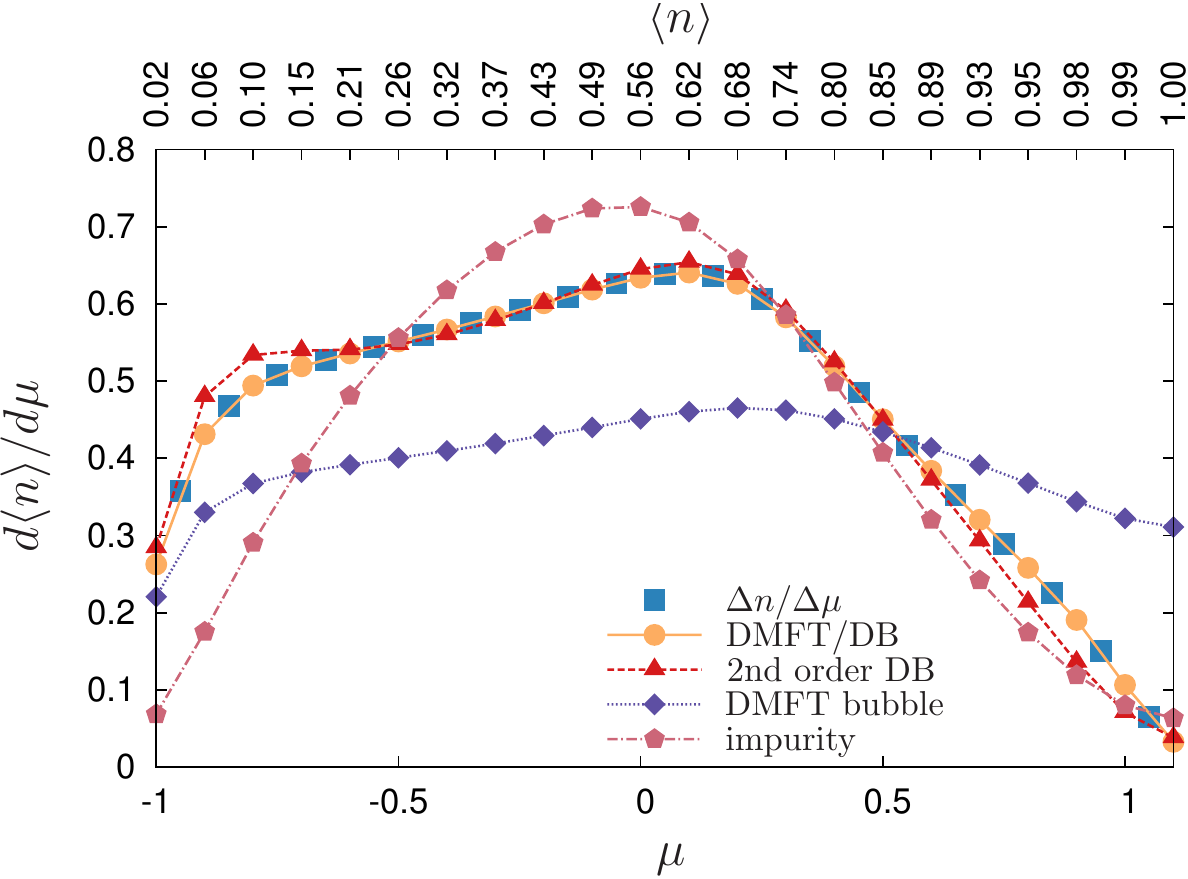}
    }
    \caption{(Color online) Compressibility $d \av{n}/d\mu$ for the Hubbard model ($V=0$), computed within various approximations and three different values of the on-site interaction. The chemical potential is shown at the bottom of each graph, the corresponding density (see Fig.~\ref{fig:nmu}) is shown at the top. Blue squares denote the response function computed by varying the chemical potential as $\Delta n/\Delta \mu$, while other results are computed from the correlation function in different approximations. Note that only $\av{n}\leq 1$ (less than half-filled) is shown, results for $\av{n}\geq 1$ can be obtained by particle-hole symmetry.
}
    \label{fig:dndmu}
  \end{figure}

Before confirming this numerically, we show the dependence of the electron density $\av{n}$ on the chemical potential $\mu$ at various interaction strengths $U$ in Fig.~\ref{fig:nmu}. At $U=0$, the exact density can be obtained directly from the non-interacting density of states.
At finite $U$ the DMFT densities are shown.
The curves approach each other in the dilute limit. For low density, the system's properties, including $\av{n}$, depend weakly on $U$ 
because the physics is determined mainly through the kinetic term.
The chemical potential is shown with the Hartree contribution $\mu_H = U\!\av{n}/2$ subtracted, so that half-filling occurs at $\mu-\mu_H=0$ for all values of $U$, due to the particle-hole symmetry of the bipartite square lattice.
The DMFT correlation function is evaluated at a fixed value of $\mu$ to find the compressibility (orange circles in Fig.~\ref{fig:dndmu}). On the other hand, the DMFT compressibility can also be obtained by numerically computing the derivative of the density with respect to $\mu$ of the curves in Fig.~\ref{fig:nmu}. Here we calculate it as the difference quotient $\Delta\! \av{n}\! / \Delta \mu$, which introduces a negligible error,  as can be judged from the plots. It is depicted by the blue squares in Fig.~\ref{fig:dndmu}.

Let us focus first on these data. As a general observation, Fig.~\ref{fig:dndmu} shows that the electron compressibility of the weakly and strongly interacting systems behaves differently as half-filling is approached. 
The results are consistent with Fermi liquid theory, i.e. the compressibility decreases with increasing interaction and decreasing spectral weight at the Fermi level.
At $U=2.35$ (cf. Fig.~\ref{fig:dndmu:U235}), the compressibility is correspondingly low. In Fig.~\ref{fig:nmu}, the $U=2.35$ graph is indeed almost flat near half-filling. Note that the compressibility does not diverge as half-filling is approached~\cite{Furukawa91,Kotliar02}, because the temperature is above the critical temperature of the metal-insulator transition.

Comparing the results for the difference quotient with DMFT/DB, we see that also numerically, the thermodynamic  consistency is apparent at all values of $U$ as expected (we recall that the DB result is the same as DMFT for $V=0$).

We further show results for the compressibility obtained from the correlation function approximated by a bubble of renormalized DMFT propagators (blue diamonds in Fig.~\ref{fig:dndmu}). 
For small interaction, the approximation performs well, as expected. Already at moderate values of the interaction however, the result deviates strongly. In particular, it does not describe the small compressibility close to half-filling. Irrespective of the value of the interaction, the result approaches the one of DMFT in the dilute limit, where vertex corrections are expected to be small.
It is further instructive to determine the compressibility from the local impurity susceptibility. This approximation neglects the fact that the impurity density depends on the hybridization, cf. Eq.~\eqref{eq:dn/dm}. While it contains local vertex corrections, it neglects nonlocal ones and the momentum dependence contained in the bubble.
As a result it performs poorly at weak interaction. For moderate and strong interaction, it is close to the response function only near half-filling, where the physics is predominantly local.

In Fig.~\ref{fig:dndmu} we examine another approximation, which in a sense interpolates between the results of the bubble and impurity susceptibility.
This approximation has been introduced in Ref.~\onlinecite{Hafermann14-2}. Without going into all details here, we mention that it is obtained by neglecting the two-particle impurity vertex and the corresponding ladder diagrams contributing to the susceptibility in the DB expression for the susceptibility [see also Appendix~\ref{app:dmftresponse}, in particular Eq.~\eqref{eq:x:ladderdb}]. It however still includes a three-leg vertex and the bubble diagram, which is renormalized through this vertex. In terms of DB perturbation theory, this diagram contributing to the dual bosonic self-energy is second order in the three-leg vertices (dual electron-boson interaction) and hence has been coined ``second-order DB''. We provide some more details on this approximation in Sec.~\ref{sec:nonlocal}.

The results of the calculation are shown as red triangles in Fig.~\ref{fig:dndmu}. At small $U$, the fermion-fermion interaction is small and neglecting the two-particle vertex works well. Furthermore, at $U=0.5$ the three-leg vertex is essentially independent of fermionic frequency~\cite{vanLoon14}, so that second-order DB is expected to behave similarly as the bubble. It agrees almost perfectly with the difference quotient. At large interaction strength, the agreement remains remarkably close.
Close to half-filling, we have seen that the local susceptibility captures the essential physics. However for quarter filling and below, we see that both vertex corrections \emph{and} the momentum dependence of the bubble are important. Neither the local susceptibility nor the bubble alone reproduce the charge response.
From the derivation of the DMFT susceptibility in Appendix~\ref{app:dmftresponse}, we see that the three-leg vertex emerges from the variation of the impurity density with respect to the hybridization $\partial \av{n}/\partial \Delta_{\nu\sigma}$ [Eq.~\eqref{eq:dn/ddelta}] and the variation of the impurity Green's function with respect to the chemical potential $\partial g_{\nu\sigma}/\partial \mu$ [Eq.~\eqref{eq:dg/dmu}]. These terms are apparently important beyond the bubble in the correlated regime.
On the other hand, neglecting the two-particle vertex corresponds to neglecting $\partial g_{\nu\sigma}/\partial\Delta_{\nu'\sigma'}$ [Eq.~\eqref{eq:dgdDelta}], i.e. the change of the impurity Green's function with respect to $\Delta_{\nu\sigma}$. 
The agreement is hence only approximate. Such an approximation has a further shortcoming, as it violates the Ward identity and charge conservation~\cite{Hafermann14-2}. 
We note however, that this approximation is computationally significantly less expensive than the full DMFT correlation function, because the two-particle impurity vertex function $\gamma$ does not have to be computed. 
In this particular case, the three-leg vertex is determined as a function of fermionic frequency only, at a single bosonic frequency $\omega_{n=0}$. This is comparable in computational cost to DMFT calculations.

Finally we note that the above discussion is relevant for the case of a local self-energy. Consistency will in general be violated if the self-energy is momentum dependent (as in dual fermion), see Ref.~\onlinecite{Otsuki14} for an example. To maintain consistency in that case requires additional diagrams in the formula for the correlation function.

\section{Nonlocal interaction}
\label{sec:nonlocal}

 \begin{figure}[t!]
    \subfloat[$U=0.5$, $V=0.2$\label{fig:dndmu:EU0500}]{%
      \includegraphics[width=.9\columnwidth]{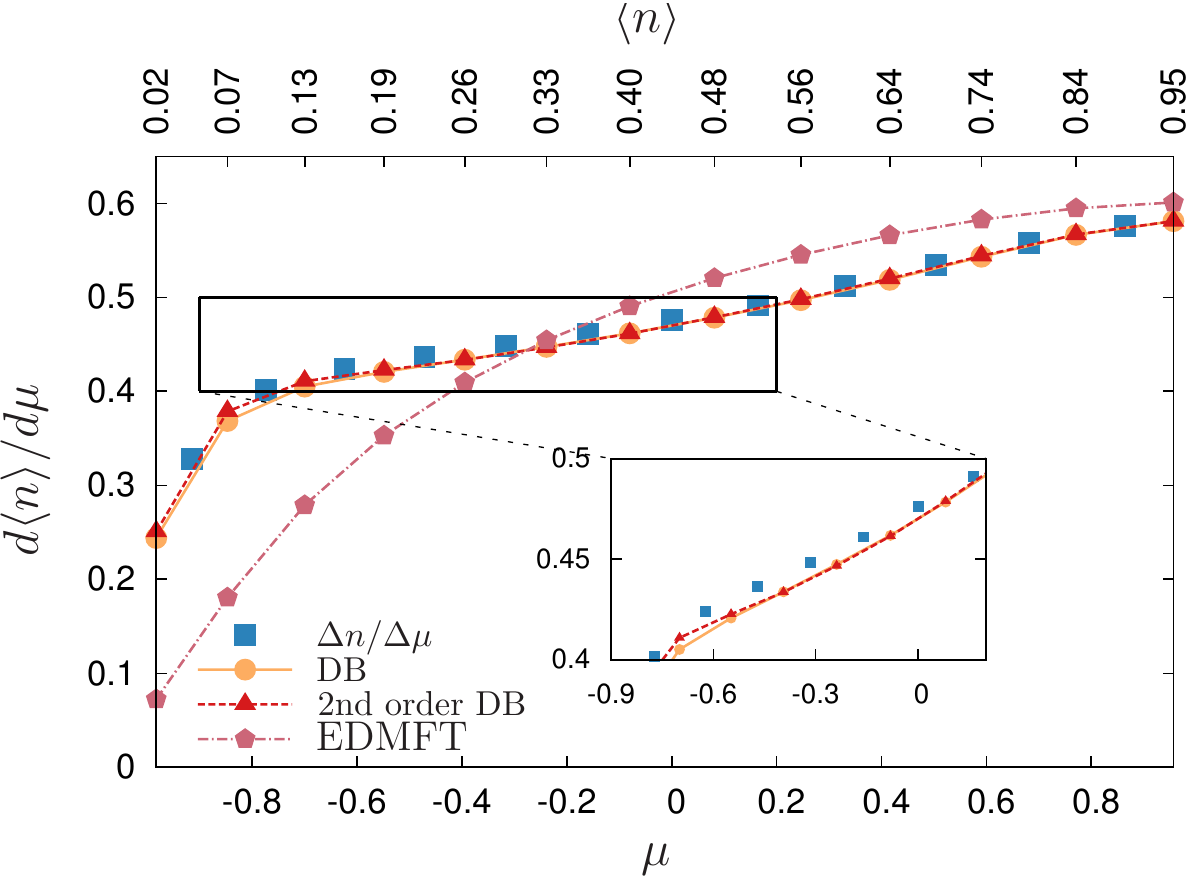}
    }
    \hfill
    \subfloat[$U=1.5$, $V=0.2$\label{fig:dndmu:EU1500}]{%
      \includegraphics[width=.9\columnwidth]{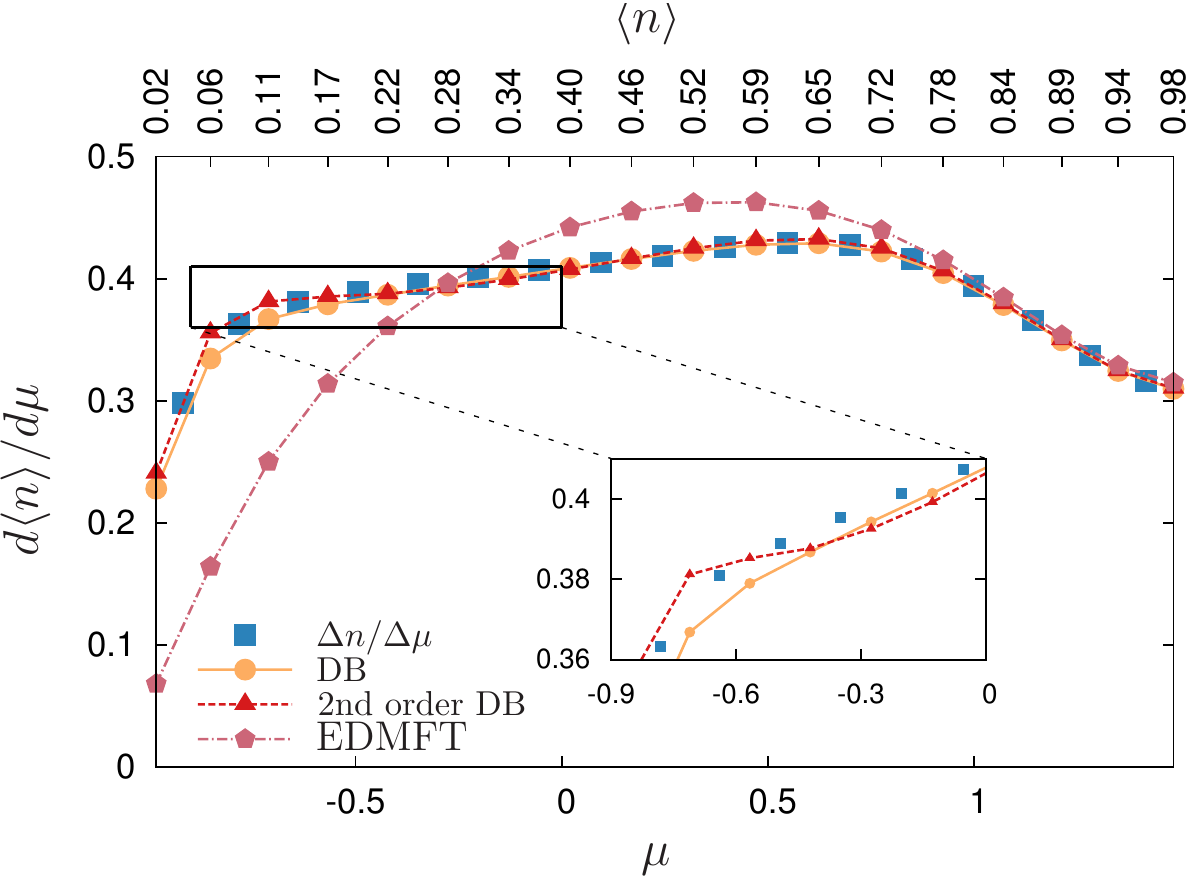}
    }
    \hfill
    \subfloat[$U=2.35$, $V=0.2$\label{fig:dndmu:EU2350}]{%
      \includegraphics[width=.9\columnwidth]{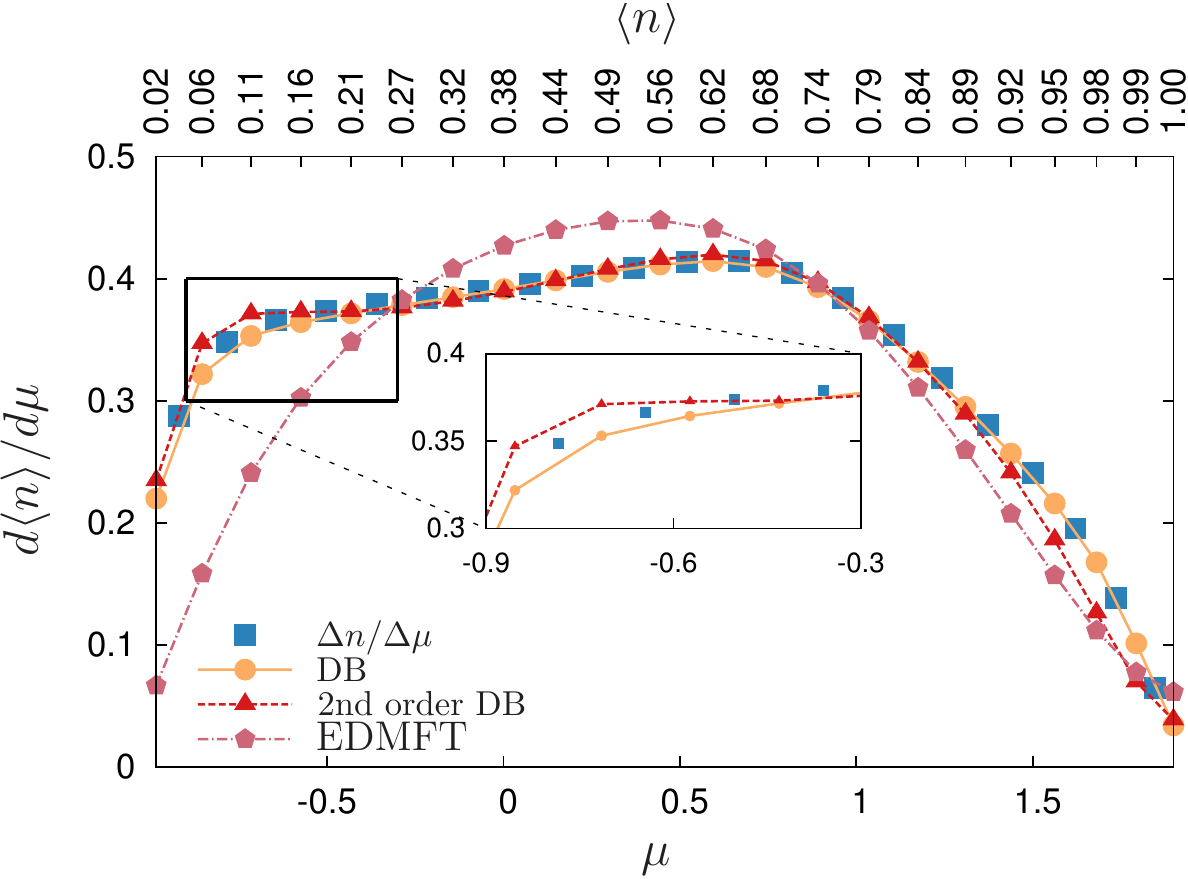}
    }
    \caption{(Color online) Compressibility as a function of chemical potential and density for finite retarded interaction. Labels are as in  Fig.~\ref{fig:dndmu}.  
    }
    \label{fig:dndmu:EDMFT}
  \end{figure}

In this section, we examine thermodynamic consistency in the extended Hubbard model within EDMFT and DB.
In EDMFT, the effect of screening due to the nonlocal interaction is accounted for through a retarded interaction in the impurity model.  This frequency dependent interaction is determined through a self-consistency condition analogous to the one for $\Delta_{\nu\sigma}$ in DMFT. We recall that the resulting impurity action reads
\begin{align}
\label{simp:edmft}
S_{\text{imp}}[c^{*},c]=&-\sum_{\nu\sigma} c^{*}_{\nu\sigma}[\inu+\mu-\Delta_{\nu\sigma}]c_{\nu\sigma}\notag\\
&+ U\sum_{\omega}n_{\omega\up}n_{-\omega\dn} + \frac{1}{2}\sum_{\omega}n_{\omega}\Lambda_{\omega} n_{-\omega}
\end{align}
and the self-consistency conditions are
\begin{align}
\sum_{\kv} G_{\kv\nu\sigma} = g_{\nu\sigma},\label{eq:scg}\\
\sum_{\qv} X_{\qv\omega} = \chi_{\omega}.
\label{eq:scchi}
\end{align}
Here $\chi_{\omega}$ denotes the impurity susceptibility and 
\begin{align}
 X^{-1}_{\qv\omega} = \chi_{\omega}^{-1} + \Lambda_{\omega} - V_{\qv} \label{x:edmft}
\end{align}
is the EDMFT charge correlation function.
In our calculations, we solve the impurity model \eqref{simp:edmft} repeatedly until the self-consistency conditions \eqref{eq:scg}, \eqref{eq:scchi} are fulfilled.

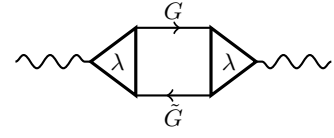
\begin{figure}[t]
\begin{tikzpicture}
    \coordinate (diagram1) at (0,0) ;

    \coordinate (leftend) at ($(diagram1)$ ) ;
    \coordinate (leftboson) at ($(leftend) + (-1.0,0)$) ;
    \coordinate (toplefttriangle) at ($(leftend) + 1.5*(0.4,0.3)$) ;
    \coordinate (bottomlefttriangle) at ($(leftend) + 1.5*(0.4,-0.3)$) ;

    \coordinate (toprighttriangle) at ($(toplefttriangle) + (1.0,0)$) ;
    \coordinate (bottomrighttriangle) at ($(toprighttriangle) + 1.5*(0,-0.6)$) ;
    \coordinate (rightend) at ($(toprighttriangle) + 1.5*(0.4,-0.3)$) ;
    \coordinate (rightboson) at ($(rightend) + (1.,0)$) ;

    \draw[very thick] (leftend) -- (toplefttriangle) -- (bottomlefttriangle) -- cycle;

    \draw[very thick] (rightend) -- (toprighttriangle) -- (bottomrighttriangle) -- cycle;

    \draw[thick,decorate,decoration=snake] (leftboson) -- (leftend) ;
    \draw[thick,decorate,decoration=snake] (rightboson) -- (rightend) ;

    \draw[thick,-<-=0.5] (bottomlefttriangle) -- (bottomrighttriangle);
    \draw[thick,-<-=0.5] (toprighttriangle) -- (toplefttriangle);

    \node[below] at ($(bottomlefttriangle)!0.5!(bottomrighttriangle)$) {$\tilde{G}$} ;
    \node[above] at ($(toplefttriangle)!0.5!(toprighttriangle)$) {$\tilde{G}$} ;
    \node at ($(leftend)+0.6*1.5*(0.4,0)$) {$\lambda$} ;    
    \node at ($(rightend)-0.6*1.5*(0.4,0)$) {$\lambda$} ;    
  \end{tikzpicture}
\caption{
Diagrammatic representation of $\tilde{\Pi}^{\text{2nd order}}_{\qv=0,\omega=0}$.
}
\label{fig:2ndorder}
\end{figure}

The DB approximation can be viewed as a diagrammatic extension of EDMFT. It allows us to incorporate additional diagrams (polarization corrections~\cite{vanLoon14-2}) into the charge susceptibility. To this end, \eqref{x:edmft} is replaced by
\begin{align}
 X^{-1}_{\qv\omega} = [\chi_{\omega}+\chi_\omega \tilde{\Pi}_{\qv\omega}\chi_\omega]^{-1} + \Lambda_{\omega} - V_{\qv}, \label{x:db}
\end{align}
where $\tilde{\Pi}_{\qv\omega}$ is dual bosonic self-energy. Note that EDMFT is recovered for $\tilde{\Pi}_{\qv\omega}\equiv 0$. We evaluate $\tilde{\Pi}$ diagrammatically. The leading diagram to $\tilde{\Pi}_{\qv\omega}$ is shown in Fig.~\ref{fig:2ndorder}. It is second order in the electron-boson interaction and we refer to this approximation as second-order DB approximation. The ladder DB is obtained by replacing one of the triangles through a renormalized triangular vertex containing ladder diagrams~\cite{vanLoon14-2}.
The explicit expressions are given by Eq.~\eqref{eq:pi:ladderdb} and Eq.~\eqref{eq:pi:2nd} respectively.

Although diagrammatic corrections to the self-energy and more elaborate self-consistency schemes are possible in DB~\cite{vanLoon14-2}, we restrict ourselves to ``one-shot''-type calculations here (see also the discussion in Sec.~\ref{sec:conclusion}). I.e., we solve the EDMFT equations, and only \emph{after} convergence evaluate the susceptibility according to \eqref{x:db} instead of \eqref{x:edmft}. As a result, the impurity quantities and in particular the Green's function and density (and hence $\Delta n/\Delta\mu$) are the same in EDMFT and DB.

In EDMFT and DB, we can determine the variation of the density with respect to the chemical potential analogously to the foregoing.
We have
\begin{align}
 \frac{d\!\av{n}}{\partial \mu} =&  \frac{\partial\!\impav{n}}{\partial \mu} +  \sum_{\nu\sigma} \frac{\partial\!\impav{n}}{\partial \Delta_{\nu\sigma}} \frac{\partial \Delta_{\nu\sigma}}{\partial \mu}\notag\\&+\sum_{\omega} \frac{\partial\!\impav{n}}{\partial \Lambda_{\omega}} \frac{\partial \Lambda_{\omega}}{\partial \mu}.\label{eq:dn/dm:edmft}
\end{align}
An additional term contributing to the compressibility emerges due to the retarded interaction $\Lambda_{\omega}$.
As discussed in the previous section, the contributions in the first line are accounted for in DB, but the one due to the retarded interaction is not.
Performing steps similar to the derivation in Appendix \ref{app:dmftresponse} shows that $\partial\!\impav{n}/\partial \Lambda_\omega \propto \impav{n_{\omega=0} n_\omega n_\omega}-\impav{n}\impav{n_\omega n_\omega}$. In fact, there will even be a four-particle correlator in the derivation, arising from $\partial\chi_{\omega}/\partial\Lambda_{\omega'}$ in the variation of the self-consistency condition (see Appendix~\ref{app:edmftresponse}).
Hence, in spite of the fact that the Kubo formula relates the response to a two-particle (lattice) correlation function, three- and four-particle impurity correlation functions appear in the expression for the response. This is due to the presence of the retarded interaction.
For most applications, accounting for three-particle or even higher-order correlations is impractical. It is therefore important to examine their contribution to observables.
There has been some discussion on the impact of three-particle interactions in the literature~\cite{Katanin13,Rohringer13}, but in numerical studies they are usually neglected~\cite{Toschi07,Rubtsov08,vanLoon14} (with the exception of Ref.~\onlinecite{Hafermann09} where the effect was found to be small).

Because of the higher-order correlation functions in the response, EDMFT and DB cannot fulfill thermodynamic consistency exactly. In order to examine to what extent they are consistent, we show the electron compressibility for the extended Hubbard model for different local and fixed nonlocal interaction strengths in Fig.~\ref{fig:dndmu:EDMFT}.
We see that the overall compressibility is smaller for finite $V$ at a given $U$ (cf. Fig.~\ref{fig:dndmu}). The agreement with $\Delta\av{n}/\Delta\mu$ is remarkably close for DB for all values of $U$. The effect of three- and four-particle correlations appears to be negligible, even close to the Mott transition.
The strongest deviations occur for relatively low filling, where they remain small (cf. insets). As before, the EDMFT susceptibility is consistent in the correlated regime close to half-filling, where the physics is predominantly local. In this regime, we have found corrections of DB perturbation theory to EDMFT to be small~\cite{vanLoon14-2}.
We have also performed a calculation at significantly lower temperature of $\beta=50$. Without nearest-neighbor interaction, this interaction strength $U=2.35$ would be very close to the metal-insulator transition. We do not find any qualitative difference to the foregoing.

\section{Conclusion and discussion}
\label{sec:conclusion}

We have assessed the thermodynamic consistency of the charge response in the (extended) Hubbard model within DMFT, EDMFT and DB.
We have proven for the Hubbard model, that DMFT yields a consistent response. 
The DMFT bubble is consistent only at small interaction. On the other hand, the local impurity susceptibility is consistent when the physics is predominantly local, i.e., at large on-site interaction and close to half-filling. Beyond these two regimes, additional diagrams are required to yield a consistent charge response. We found that dominant contributions are retained even when neglecting nonlocal vertex corrections from the two-particle vertex (this amounts to a second-order approximation in terms of dual variables).
From these results we conclude that the most important ingredients for maintaining a consistent response is the momentum dependence at low interaction, local vertex corrections at strong interaction and their interplay at intermediate interaction, whereby a renormalized electron-boson coupling given by the impurity three-leg vertex is essential.

In the extended Hubbard model with nearest-neighbor interaction, we have shown that the response is determined not only by two- but also by three- and four-particle correlations. This is a consequence of the self-consistent retarded interaction included in the impurity model. The DB approximation, which neglects associated diagrams, however, yields a response which is consistent to good approximation.
Also here, non-local long-range vertex corrections appear to have a small effect. At least in this aspect, the effect of third- and fourth-order correlation functions is negligible, which is important in practice. EDMFT is, in general, not consistent. It is consistent when the physics is essentially local. We found that the most important contribution beyond DMFT to correct this deficit is a second-order diagram in the renormalized triangular vertex, which effectively includes momentum dependence into the EDMFT polarization.

The above observations are valid in the case when no diagrammatic corrections to the fermionic self-energy are taken into account, i.e. it remains local and equal to the impurity self-energy. For a non-local self-energy, maintaining consistency requires additional contributions to the correlation function.

\acknowledgments
The authors acknowledge helpful discussions with Andrey Katanin and thank Remko Logemann and Patrick Graman for their comments on the manuscript. E.G.C.P. v. L. and M.I.K. acknowledge support from ERC Advanced Grant 338957 FEMTO/NANO. A.L. is supported by the DFG-FOR1346 program.
Our DB implementation employs an extended version of the CT-HYB solver of Ref.~\onlinecite{Hafermann13}. The solver and the DB implementation are based on the ALPS libaries~\cite{ALPS2}.

\appendix

\section{Linear response formalism}
\label{app:kubo}

The particle number and the density response can be determined from the partition function or from correlation functions. Starting from the Hamiltonian \eqref{eq:hmlt}, the imaginary time path integral formalism~\cite{NegeleOrland} allows all observables to be calculated using the partition function
$
Z = \int D[c^*,c] \exp(-S[c^*,c])
$, with the Euclidean action $S[c^*,c]= \int_0^\beta d\tau [c^* \partial /\partial \tau~c + H]$.
The average number of particles per site, $\av{n}$, is obtained by deriving with respect to the chemical potential,
\begin{align}
 \av{n} 
 &= \frac{1}{N}\sum_{j} \frac{1}{Z} \int D[c^*,c]\; n_j \; \exp(-S[c^*,c]) \notag\\
 &= \frac{1}{\beta N} \frac{1}{Z} \frac{\partial Z}{\partial \mu},  \label{eq:nfromz}
\end{align}
where $j$ is a site index and $N$ is the total number of sites. Similarly, the compressibility is
\begin{align}
 \frac{d\!\av{n}}{d\mu} &= 
 \frac{1}{\beta N} \frac{\partial}{\partial \mu}\left(\frac{1}{Z} \frac{\partial}{\partial \mu} Z\right) \notag \\
 &=\frac{1}{\beta N} \left[\frac{1}{Z}\frac{\partial^2 Z}{\partial \mu^2} - \left(\frac{1}{Z}\frac{\partial Z}{\partial \mu}\right)^2 \right] \notag \\ 
 &= \frac{1}{\beta N} \sum_{jk} \int d\tau_1 d\tau_2 \av{n_j(\tau_1) n_k(\tau_2)} - \av{n_j(\tau_1)}\av{n_k(\tau_2)} \notag \\
 &= -X_{q=0,\omega=0},
\end{align}
where $X_{q=0,\omega_n=0}$ is to be understood as the static uniform charge correlation function, where the limit $\omega\rightarrow 0$ has been taken before the limit $q\rightarrow 0$. These limits do not commute in general~\cite{Hafermann14-2}.

\section{DMFT response formula}
\label{app:dmftresponse}

In this Appendix, we show that the DMFT and DB response in the Hubbard model are consistent.
To this end, we evaluate the individual terms in the expression \eqref{eq:dn/dm} for the electron compressibility,
\begin{align}
 \frac{d\!\av{n}}{d \mu} &=  \frac{\partial\!\impav{n}}{\partial \mu} +  \sum_{\nu\sigma} \frac{\partial\! \impav{n}}{\partial \Delta_{\nu\sigma}} \frac{\partial \Delta_{\nu\sigma}}{\partial \mu}.
\end{align}
This is done by rewriting derivatives of impurity quantities in terms of impurity correlation functions. Since the impurity problem \eqref{simp:dmft} is solved (numerically) exactly, it is thermodynamically consistent and this rewriting is valid.
The variation of the hybridization with respect to the chemical potential depends on the self-consistency condition \eqref{eq:sc}, which includes the effect of the lattice and determines the derivatives. 
For compactness, all factors of $\beta$ are suppressed in this derivation and the notation of Ref.~\onlinecite{vanLoon14-2} is adopted for the impurity correlation functions. 

All impurity expectation values are obtained from the impurity partition function
\begin{align}
 \mathcal{Z} = \int D[c^*,c] \exp(-S_{\text{imp}}).
\end{align}
For example, the impurity density is 
\begin{align}
\impav{n} &= 1/\mathcal{Z} \int D[c^*,c] n \exp(-S_{\text{imp}}) \notag \\
&= 1/\mathcal{Z} \int D[c^*,c] \frac{\partial}{\partial \mu} \exp(-S_{\text{imp}}).
\end{align}
This leads to an impurity compressibility
\begin{align}
\frac{\partial\!\impav{n}}{\partial \mu}  
&= 1/\mathcal{Z} \int D[c^*,c] \frac{\partial^2}{\partial \mu^2} \exp(-S_{\text{imp}}) \notag \\
&\phantom{=}-\left(1/\mathcal{Z} \int D[c^*,c] \frac{\partial}{\partial \mu} \exp(-S_{\text{imp}})\right)^2 \notag \\
&= \impav{nn}_{\omega=0} - \impav{n}\impav{n} \notag \\
&\teL - \chi_{\omega=0}
\end{align}

The variation of the density with respect to the hybridization yields
\begin{align}
\frac{\partial\!\impav{n}}{\partial \Delta_{\nu\sigma}}  
=& 1/\mathcal{Z} \int D[c^*,c] n\frac{\partial}{\partial \Delta_{\nu\sigma}} \exp(-S_{\text{imp}}) \notag \\
&-\left(1/\mathcal{Z} \int D[c^*,c] n \exp(-S_{\text{imp}})\right) \notag \\
&\times \left(1/\mathcal{Z} \int D[c^*,c] \frac{\partial}{\partial \Delta_{\nu\sigma}} \exp(-S_{\text{imp}})\right) \notag \\
=& -1/\mathcal{Z} \int D[c^*,c] n c^{*}_{\nu\sigma}c^{\phantom{*}}_{\nu\sigma} \exp(-S_{\text{imp}}) \notag \\
&+\impav{n}\left(1/\mathcal{Z} \int D[c^*,c] c^{*}_{\nu\sigma}c^{\phantom{*}}_{\nu\sigma} \exp(-S_{\text{imp}})\right) \notag \\
=& - \impav{n c^{*}_{\nu\sigma}c^{\phantom{*}}_{\nu\sigma}}+\impav{n}\impav{c^{*}_{\nu\sigma}c^{\phantom{*}}_{\nu\sigma}} \notag \\
\teL& - \lambda_{\nu\sigma,\omega=0}\chi_{\omega=0}g_{\nu\sigma}g_{\nu\sigma},
\label{eq:dn/ddelta}
\end{align}
where $\lambda_{\nu\sigma,\omega}$ is the impurity three-leg vertex function~\cite{vanLoon14} and $g_{\nu\sigma} = -\impav{c^{\phantom{*}}_{\nu\sigma}c^{*}_{\nu\sigma}}$ is the impurity Green's function. 
Variation of the impurity Green's function with respect to the chemical potential gives 
\begin{align}
 \frac{\partial g_{\nu\sigma}}{\partial \mu} =& -1/\mathcal{Z} \int D[c^*,c] c^{\phantom{*}}_{\nu\sigma}c^{*}_{\nu\sigma}\frac{\partial}{\partial \mu} \exp(-S_{\text{imp}}) \notag \\
&+\left(1/\mathcal{Z} \int D[c^*,c] c^{\phantom{*}}_{\nu\sigma}c^{*}_{\nu\sigma} \exp(-S_{\text{imp}})\right)\notag \\
&\times \left(1/\mathcal{Z} \int D[c^*,c] \frac{\partial}{\partial \mu} \exp(-S_{\text{imp}})\right) \notag\\
=& -1/\mathcal{Z} \int D[c^*,c] c^{\phantom{*}}_{\nu\sigma}c^{*}_{\nu\sigma}n_{\omega=0} \exp(-S_{\text{imp}}) \notag \\
&+\left(1/\mathcal{Z} \int D[c^*,c] c^{\phantom{*}}_{\nu\sigma}c^{*}_{\nu\sigma} \exp(-S_{\text{imp}})\right)\notag \\
&\times \left(1/\mathcal{Z} \int D[c^*,c] n_{\omega=0} \exp(-S_{\text{imp}})\right) \notag\\
=&-\impav{c^{\phantom{*}}_{\nu\sigma}c^{*}_{\nu\sigma} n_{\omega=0}} + \impav{c^{\phantom{*}}_{\nu\sigma}c^{*}_{\nu\sigma}} \impav{n} \notag\\
\label{eq:dg/dmu}
=& \chi_{\omega=0}g_{\nu\sigma}g_{\nu\sigma}\lambda_{\nu\sigma,\omega=0}
\end{align}
The impurity Green's function $g_{\nu\sigma}$ depends on the hybridization $\Delta_{\nu'\sigma'}$ even when the frequency and spin indices are different:
\begin{align}
 -\frac{\partial g_{\nu\sigma}}{\partial \Delta_{\nu'\sigma'}} =& 1/\mathcal{Z} \int D[c^*,c] c^{\phantom{*}}_{\nu\sigma}c^{*}_{\nu\sigma}\frac{\partial}{\partial \Delta_{\nu'\sigma'}} \exp(-S_{\text{imp}}) \notag \\
&-\left(1/\mathcal{Z} \int D[c^*,c] c^{\phantom{*}}_{\nu\sigma}c^{*}_{\nu\sigma} \exp(-S_{\text{imp}})\right)\notag \\
&\times \left(1/\mathcal{Z} \int D[c^*,c] \frac{\partial}{\partial \Delta_{\nu'\sigma'}} \exp(-S_{\text{imp}})\right) \notag\\
=& -1/\mathcal{Z} \int D[c^*,c] c^{\phantom{*}}_{\nu\sigma}c^{*}_{\nu\sigma} c^{*}_{\nu'\sigma'}c^{\phantom{*}}_{\nu'\sigma'} \exp(-S_{\text{imp}}) \notag \\
&+\left(1/\mathcal{Z} \int D[c^*,c] c^{\phantom{*}}_{\nu\sigma}c^{*}_{\nu\sigma} \exp(-S_{\text{imp}})\right)\notag \\
&\times \left(1/\mathcal{Z} \int D[c^*,c] c^{*}_{\nu'\sigma'}c^{\phantom{*}}_{\nu'\sigma'} \exp(-S_{\text{imp}})\right) \notag\\
=&\impav{c^{\phantom{*}}_{\nu\sigma}c^{*}_{\nu\sigma} c^{\phantom{*}}_{\nu'\sigma'}c^{*}_{\nu'\sigma'}} \notag \\
&-  \impav{c^{\phantom{*}}_{\nu\sigma}c^{*}_{\nu\sigma}} \impav{c^{\phantom{*}}_{\nu'\sigma'}c^{*}_{\nu'\sigma'}} \notag\\
=& \gamma_{\nu\nu'\sigma\sigma',\omega=0} \times g_{\nu\sigma}g_{\nu\sigma}g_{\nu'\sigma'}g_{\nu'\sigma'} \notag \\ &-\delta_{\nu\nu'}\delta_{\sigma\sigma'}g_{\nu\sigma}g_{\nu\sigma},
\label{eq:dgdDelta}
\end{align}
where we have used the definition of the two-particle impurity vertex function (see, e.g., Ref~\onlinecite{vanLoon14}). In the non-interacting limit, $\gamma=0$ and $g_{\nu\sigma}^{(0)}=1/(\inu -\Delta_{\nu\sigma})$. Taking the derivative $-\partial g^{(0)}_{\nu\sigma}/\partial \Delta_{\nu'\sigma'}$ indeed gives $-\delta_{\nu\nu'}\delta_{\sigma\sigma'}g^{(0)}_{\nu\sigma}g^{(0)}_{\nu\sigma}$ as in \eqref{eq:dgdDelta}.

The DMFT self-consistency condition determines the hybridization function $\Delta_{\nu\sigma}$ such that the local Green's function is equal to the impurity Green's function for all Matsubara frequencies $\nu$:
\begin{align}
 0=&g_{\nu\sigma} - \sum_{\kv} G_{\nu\sigma\kv}\notag\\
  =&g_{\nu\sigma} - \sum_{\kv} \frac{1}{g_{\nu\sigma}^{-1}+\Delta_{\nu\sigma}-t_\kv} \notag\\
 =&f_{\nu\sigma}(\mu,\{\Delta_{\nu'\sigma'}\})
\end{align}
with a factor $1/N$ implied in the sum over $\kv$. Here we have introduced $f_{\nu\sigma}(\mu,\{\Delta_{\nu'\sigma'}\})$ for notational convenience. 
The partial derivative of the self-consistency condition with respect to the chemical potential at fixed $\{\Delta_{\nu'\sigma'}\}$ is
\begin{align}
 \frac{\partial f_{\nu\sigma}}{\partial \mu}\bigg|_{\{\Delta_{\nu'\sigma'} \}} 
 &= \frac{\partial g_{\nu\sigma}}{\partial \mu} + \sum_{\kv} \frac{1}{(g_{\nu\sigma}^{-1}+\Delta_{\nu\sigma}-t_\kv)^2} \frac{\partial g_{\nu\sigma}^{-1}}{\partial \mu} \notag \\
 &= \frac{\partial g_{\nu\sigma}}{\partial \mu} \left(1 - g^{-2}_{\nu\sigma} \sum_{\kv} G^2_{\nu\sigma\kv} \right) \notag \\
 &=  \frac{\partial g_{\nu\sigma}}{\partial \mu} g^{-2}_{\nu\sigma} \left(g^{2}_{\nu\sigma} - \sum_{\kv} G^2_{\nu\sigma\kv} \right) \notag \\
 &=  -\frac{\partial g_{\nu\sigma}}{\partial \mu} g^{-2}_{\nu\sigma} \sum_{\kv} \tilde{G}^2_{\nu\sigma\kv}.
\end{align}
Here we have introduced the dual Green's function $\tilde{G} = G-g$ and we made use of the property
\begin{align}
 \sum_{\kv} G^2_{\nu\sigma\kv} =& \sum_{\kv} (\tilde{G}_{\nu\sigma\kv}+g_{\nu\sigma})(\tilde{G}_{\nu\sigma\kv}+g_{\nu\sigma}) \notag\\
 =& \sum_{\kv} \tilde{G}^2_{\nu\sigma\kv}+ 2 g_{\nu\sigma} \sum_{\kv} \tilde{G}_{\nu\sigma\kv}+ g^2_{\nu\sigma} \notag\\
=& \sum_{\kv} \tilde{G}^2_{\nu\sigma\kv}+ g^2_{\nu\sigma}. \label{eq:bubbles}
\end{align}
This holds since the DMFT self-consistency condition \eqref{eq:sc} implies $\sum_\kv \tilde{G}_{\nu\sigma\kv} =0$. 
The dependence on $\Delta_{\nu\sigma}$ requires some attention. The Green's function $G_{\nu\sigma\kv}$ depends on $\Delta_{\nu\sigma}$ explicitly, but also implicitly through $g_{\nu\sigma}$. As shown in \eqref{eq:dgdDelta}, $g_{\nu\sigma}$ depends on $\Delta_{\nu'\sigma'}$ also for \emph{different} Matsubara frequencies. This means that all the self-consistency conditions depend on all components of $\Delta_{\nu\sigma}$, so the derivative of the self-consistency condition will be a matrix equation in frequency space:
\begin{widetext}
\begin{align}
 \frac{\partial f_{\nu\sigma}}{\partial \Delta_{\nu'\sigma'}}\bigg|_{\mu} 
 &= \frac{\partial g_{\nu\sigma}}{\partial \Delta_{\nu'\sigma'}} + \sum_{\kv} \frac{1}{(g_{\nu\sigma}^{-1}+\Delta_{\nu\sigma}-t_\kv)^2} \left(\frac{\partial g_{\nu\sigma}^{-1}}{\partial \Delta_{\nu'\sigma'}}+\delta_{\nu\nu'}\delta_{\sigma\sigma'}\right) \notag\\
 &= \frac{\partial g_{\nu\sigma}}{\partial \Delta_{\nu'\sigma'}} \left(1 - g^{-2}_{\nu\sigma} \sum_{\kv} G^2_{\nu\sigma\kv} \right) + \delta_{\nu\nu'}\delta_{\sigma\sigma'}\sum_{\kv} G^2_{\nu\sigma\kv} \notag\\
 &= -g^{-2}_{\nu\sigma}\frac{\partial g_{\nu\sigma}}{\partial \Delta_{\nu'\sigma'}} \sum_{\kv} \tilde{G}^2_{\nu\sigma\kv} + \delta_{\nu\nu'}\delta_{\sigma\sigma'}\sum_{\kv} G^2_{\nu\sigma\kv}. 
\end{align}
Collecting previous results, the variation of the self-consistency condition gives
\begin{align}
 0=& \frac{\partial f_{\nu'\sigma'}}{\partial \mu} + \sum_{\nu\sigma} \frac{\partial f_{\nu'\sigma'}}{\partial \Delta_{\nu\sigma}} \frac{\partial \Delta_{\nu\sigma}}{\partial \mu} \notag\\
 \frac{\partial \Delta_{\nu\sigma}}{\partial \mu} =& - \sum_{\nu'\sigma'} \left[ \frac{\partial f_{\nu'\sigma'}}{\partial \Delta_{\nu\sigma}} \right]^{-1} \frac{\partial f_{\nu'}}{\partial \mu} \notag\\
=& \sum_{\nu'\sigma'} \left[ -g^{-2}_{\nu'\sigma'}\frac{\partial g_{\nu'\sigma'}}{\partial \Delta_{\nu\sigma}} \sum_{\kv} \tilde{G}^2_{\nu'\sigma'\kv} + \delta_{\nu\nu'}\delta_{\sigma\sigma'}\sum_{\kv} G^2_{\nu\sigma\kv} \right]^{-1} 
\left(\frac{\partial g_{\nu'\sigma'}}{\partial \mu} g^{-2}_{\nu'\sigma'} \sum_{\kv} \tilde{G}^2_{\nu'\sigma'\kv}\right) \notag\\
=& \sum_{\nu'\sigma'} \left[ -\frac{\partial g_{\nu'\sigma'}}{\partial \Delta_{\nu\sigma}} \sum_{\kv} \tilde{G}^2_{\nu'\sigma'\kv} + \delta_{\nu\nu'}\delta_{\sigma\sigma'}g^{2}_{\nu'\sigma'}\sum_{\kv} G^2_{\nu'\sigma'\kv} \right]^{-1} \frac{\partial g_{\nu'\sigma'}}{\partial \mu} \sum_{\kv} \tilde{G}^2_{\nu'\sigma'\kv} \notag\\
\overset{\eqref{eq:bubbles}}{=}& \sum_{\nu'\sigma'} \left[ \left[\delta_{\nu\nu'}\delta_{\sigma\sigma'}g^{2}_{\nu'\sigma'}-\frac{\partial g_{\nu'\sigma'}}{\partial \Delta_{\nu\sigma}}\right] \sum_{\kv} \tilde{G}^2_{\nu'\sigma'\kv} + \delta_{\nu\nu'}\delta_{\sigma\sigma'}g^{4}_{\nu'\sigma'} \right]^{-1}
\frac{\partial g_{\nu'\sigma'}}{\partial \mu} \sum_{\kv} \tilde{G}^2_{\nu'\sigma'\kv} \notag\\
\overset{\eqref{eq:dgdDelta}}{=}& 
\sum_{\nu'\sigma'} \left[ \gamma_{\nu'\nu\sigma'\sigma,\omega=0} \sum_{\kv} \tilde{G}^2_{\nu'\sigma'\kv} + \delta_{\nu\nu'}\delta_{\sigma\sigma'} \right]^{-1}
g^{-4}_{\nu'\sigma'}\frac{\partial g_{\nu'\sigma'}}{\partial \mu} \sum_{\kv} \tilde{G}^2_{\nu'\sigma'\kv} \label{eq:variation-sc}
\end{align}
Here $\left[ \frac{\partial f_{\nu'\sigma'}}{\partial \Delta_{\nu\sigma}} \right]^{-1}$ and all other inverses should be understood as a matrix inversion in spin- and frequency space.
Finally, the lattice compressibility is obtained by combining the above results:
\begin{align}
 X_{\qv=0,\omega=0} =& -\frac{d \av{n}}{d \mu} = -\frac{\partial \impav{n}}{\partial \mu} - \sum_{\nu\sigma} \frac{\partial \impav{n}}{\partial \Delta_{\nu\sigma}} \frac{\partial \Delta_{\nu\sigma}}{\partial \mu} \notag \\
 =& \chi_{\omega=0}  + \sum_{\nu\nu'\sigma\sigma'} 
 \chi_{\omega=0}\lambda_{\nu\sigma,\omega=0}
 \sum_{\kv} \tilde{G}^2_{\nu\sigma\kv} \left[ \gamma_{\nu\nu\sigma\sigma,\omega=0} \sum_{\kv} \tilde{G}^2_{\nu\sigma\kv} + \delta_{\nu\nu'}\delta_{\sigma\sigma'} \right]^{-1}
 \lambda_{\nu'\sigma',\omega=0}\chi_{\omega=0} \label{eq:x:ladderdb}
\end{align}
This is equal to the DB expression~\cite{Rubtsov12,vanLoon14,vanLoon14-2,Hafermann14-2} for $V=0$ and $\Lambda=0$, i.e., $X_{\qv\omega} = \chi_{\omega}  + \chi_{\omega}\tilde{\Pi}_{\qv\omega}\chi_{\omega} $, with dual bosonic self-energy
\begin{align}
 \tilde{\Pi}_{\qv=0,\omega=0} &=  \sum_{\nu\nu'\sigma\sigma'} \lambda_{\nu\sigma,\omega=0}\sum_{\kv} \tilde{G}^2_{\nu\sigma\kv}
 \left[ \gamma_{\nu'\nu\sigma'\sigma,\omega=0} \sum_{\kv} \tilde{G}^2_{\nu'\sigma'\kv} + \delta_{\nu\nu'}\delta_{\sigma\sigma'} \right]^{-1}  \lambda_{\nu'\sigma',\omega=0}.\label{eq:pi:ladderdb}
\end{align}
The ladder DB correlation function has been proven to be equal to the DMFT correlation function~\cite{Hafermann14-2}, completing the proof of the consistency of the DMFT response function.

\end{widetext}

\section{Second-order Dual Boson}

Eq.~\eqref{eq:pi:ladderdb} is the dual bosonic self-energy in the ladder approximation.
In this paper we also use a simplier approximation, which was also introduced in Ref.~\onlinecite{vanLoon14-2}.
It is obtained by neglecting the vertex $\gamma$ in \eqref{eq:pi:ladderdb}, so that
\begin{align}
 \left[ \gamma_{\nu'\nu\sigma'\sigma,\omega=0} \sum_{\kv} \tilde{G}^2_{\nu'\sigma'\kv} + \delta_{\nu\nu'}\delta_{\sigma\sigma'} \right]^{-1} \rightarrow \delta_{\nu\nu'}\delta_{\sigma\sigma'}.
\end{align}
The resulting Feynman diagram (see Fig.~\ref{fig:2ndorder}) corresponds to a second-order approximation to the dual bosonic self-energy and reads
\begin{align}
 \tilde{\Pi}^{\text{2nd order}}_{\qv=0,\omega=0} &=  \sum_{\nu\sigma} \lambda_{\nu\sigma,\omega=0}\sum_{\kv} \tilde{G}^2_{\nu\sigma\kv}
 \lambda_{\nu\sigma,\omega=0}.\label{eq:pi:2nd}
\end{align}

To get a feeling for what this approximation means, we can look back to the derivation of Eq.~\eqref{eq:pi:ladderdb} and see that $\gamma$ appeared in Eq.~\eqref{eq:dgdDelta} from the dependence of $g_{\nu\sigma}$ on $\Delta_{\nu'\sigma'}$.
Second-order DB does not fully take this dependence into account, causing some degree of thermodynamic inconsistency.
On the other hand, second-order DB does contain the three-leg vertices $\lambda$, coming from both the variation of the impurity density with respect to the hybridization and the variation of the impurity Green's function with respect to the chemical potential.

\section{EDMFT response}
\label{app:edmftresponse}

As shown in \eqref{eq:dn/dm:edmft}, additional contributions to the electron compressibility appear in EDMFT. The self-consistent determination of the impurity retarded interaction $\Lambda_\omega$ leads to additional terms compared to the result of Appendix \ref{app:dmftresponse}. Here we sketch the derivation and how a three- and a four-particle correlation function arise. The three-particle correlation function is the result of the derivative $\partial\! \impav{n}/\partial\Lambda_{\omega}\propto \impav{n_{\omega=0}n_\omega n_{-\omega}}-\impav{n}\impav{nn}_{\omega}$.
The variation of the retarded interaction $\Lambda_\omega$ with respect to $\mu$ is determined similarly compared to that of $\Delta_{\nu\sigma}$ in \eqref{eq:variation-sc}. The equivalent of $\partial g_{\nu\sigma}/\partial \Delta_{\nu'\sigma'}$ is, for $\omega\neq 0$,
\begin{align}
 \frac{\partial \chi_\omega}{\partial \Lambda_{\omega'}} 
 =&
 -\frac{\partial}{\partial \Lambda_{\omega'}} \frac{1}{\mathcal{Z}} \int D[c^*,c] n_\omega n_\omega \exp(-S[c^*,c]) \notag \\
 =& 
 \frac{1}{\mathcal{Z}}\int D[c^*,c] n_\omega n_\omega n_{\omega'} n_{\omega'} \exp(-S[c^*,c]) \notag \\
 &-\frac{1}{\mathcal{Z}}\int D[c^*,c] n_\omega n_\omega \exp(-S[c^*,c])\notag \\
 &\times \frac{1}{\mathcal{Z}}\int D[c^*,c] n_{\omega'}n_{\omega'} \exp(-S[c^*,c]) \notag \\
 =& \impav{nnnn}_{\omega\omega'} - \impav{nn}_{\omega}\impav{nn}_{\omega'},
\end{align}
i.e., it gives rise to a four-particle correlator. For $\omega=0$, additional terms arise, since $\chi_\omega = \impav{nn}_\omega - \delta_{\omega} \impav{n}\impav{n}$.
There are further differences in the derivation. The EDMFT self-consistency conditions \eqref{eq:scg} and \eqref{eq:scchi} depend on both $\Delta_{\nu\sigma}$ \emph{and} $\Lambda_{\omega}$ via the impurity expectation values. This means that the formula for the variation of $\Delta_{\nu\sigma}$ \eqref{eq:variation-sc} will contain contributions from $\Lambda_{\nu\sigma}$ and vice versa. Then $\partial \Delta_{\nu\sigma}/\partial \mu$ is obtained by inverting 
\begin{align}
  0=& \frac{\partial f_{\nu'\sigma'}}{\partial \mu} + \sum_{\nu\sigma} \frac{\partial f_{\nu'\sigma'}}{\partial \Delta_{\nu\sigma}} \frac{\partial \Delta_{\nu\sigma}}{\partial \mu} + \sum_{\omega} \frac{\partial f_{\nu'\sigma'}}{\partial \Lambda_{\omega}} \frac{\partial \Lambda_{\omega}}{\partial \mu}.
\end{align}

\bibliography{main}

\end{document}